\documentclass{elsarticle}
\usepackage{epsfig,amsmath,amssymb,mathtools,amscd}
\usepackage[active]{srcltx}
\usepackage{subfigure}
\usepackage{enumerate}

\def\eg{{e.~g.} }
\mathchardef\minus="002D

\def\<{\langle}\def\>{\rangle}

\def\L2{{\mathcal L}_2}

\def\d#1 {\mathop{\mathrm{d}#1}}
\def\df#1#2 {\frac{\mathop{\mathrm{d}#1}}{#2}}

\begin{document}
\title{Insights Into Quantitative Biology: analysis of cellular
  adaptation}
\author{V.~Agoni}\ead{valentina.agoni@unipv.it} \address{Dipartimento
  di Medicina Molecolare dell'Universit\`a degli Studi di Pavia, via
  Forlanini 6, 27100 Pavia}

\begin{abstract}
  In the last years many powerful techniques have emerged to measure
  protein interactions as well as gene expression. Many progresses
  have been done since the introduction of these techniques but not
  toward quantitative analysis of data.  In this paper we show how to
  study cellular adaptation and how to detect cellular
  subpopulations. Moreover we go deeper in analyzing signal
  transduction pathways dynamics.
\end{abstract}

\begin{keyword} Real Time PCR \sep Signal transduction pathways
  dynamics \sep Cellular subpopulations \end{keyword}

\maketitle  
\section{Introduction}

To study protein interaction innovative technique are available as
isothermal titration calorimetry (ITC), differential scanning
calorimetry (DSC) and surface plasmon resonance (SPR). Although
western blot remains the more useful method to access protein levels
in the cell.  Consider that although many powerful and sensitive
techniques we have, at the end usually we reduce to express the
results with “arbitrary units” making very difficult for example to
compare articles.  Moreover if we want to investigate the reaction to
a stimulus, in the data analysis process we should consider some
parameters: Michaelis-Menten constant, antibody affinity and signal
transduction pathways amplification.  In this paper we show how to
study cellular adaptation taking into account all these parameters.

At the present days in many laboratories it is also possible to get a
quantitative analysis on gene expression using Real Time PCR. Real
Time PCR permit to compare different articles, moreover its
sensitivity is very high but we can not easily identify small cellular
subpopulations. Droplet Digital PCR solves the problem by single
events amplification.  However in this article we are going to show
that it can be possible to detect a subpopulation also using Real Time
PCR and a small amount of cDNA.

Currently the state of art in biological studies is to just indicate
trends (obtained with quantitative methods of course) with no mention
to any attempt to quantify cellular adaptation or evolution. Nor it is
contemplate the option that some tissues are themselves heterogeneous
populations (as in the case of muscles), in other cases cells react in
different ways to an injury.  Our purpose is to consider all these
tasks to give a quantitative analysis using existing biological
techniques.

\section {Quantifying cellular adaptation}
Since the pioneering work of Michaelis and Menten
\cite{michaelis1913kinetics}, the understanding of enzyme kinetics
gained an increasing interest from the community. Recent advances in
room-temperature single-molecule fluorescence studies have allowed
very precise measurements \cite{english2005ever}. On the other hand,
peptide arrays resulted to be a key technology for deciphering enzyme
function \cite{thiele2011deciphering}. Quantitative analysis of enzyme
kinetics have been developed by Se-Hui Jung
\cite{jung2012quantitative} using fluorescence-conjugated peptide
arrays, a surface concentration-based assay with solid phase. This
assay was successfully applied for calculating the Michaelis-Menten
constant ($K_M$), defined as the substrate concentration at which the
enzyme works at the half of its maximal velocity. In addition, in the
last years, many powerful techniques have emerged to study protein
interactions along with typical parameters involved in these
processes. Isothermal Titration Calorimetry (ITC) is the gold standard
for measuring binding constants ($K_B$), reaction stoichiometry ($n$),
enthalpy ($\Delta H$) and entropy ($\Delta S$), Differential Scanning
Calorimetry (DSC) is designed to study thermal stability and finally
Surface Plasmon Resonance (SPR) allows the determination of
concentration and binding affinity
\cite{hahnefeld2004determination}. These techniques could also be
applied to antibodies \cite{linnebacher2012clonality,kikuchi2005determination}.

In order to investigate the reaction to a stimulus, Western Blot is
considered the more useful method to access protein
concentration. However, in the data analysis process usually it is not
taken into account that the relative abundance of an enzyme could
affect its kinetic properties. 

In a typical situation the variation of concentration of some enzymes
with respect to control is considered to study a particular aspect,
e.g. metabolic adaptation and degradation systems.  We argue that in
some cases it is not possible to coherently interpret the results
considering only enzymes variations without including in the analysis
the relation between them and the enzymes kinetic properties, in
particular their Michaelis-Menten constant. 

\begin{figure}[h!]
\centering
\includegraphics[width=.8\textwidth]{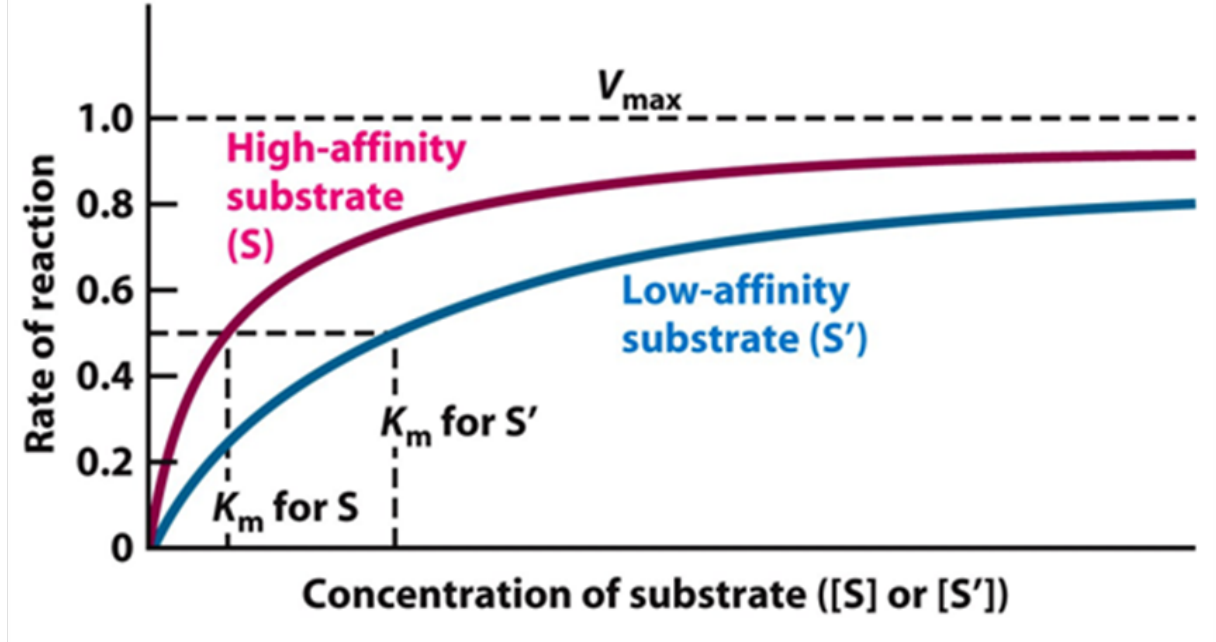}
\caption{Reaction rate vs concentration of two enzymes $A$ and
  $B$, respectively the bottom and the top one, having different
  efficiency. In the figure $S$ denotes the substrate
  concentration. One can see that the higher is the $K_M$ of the
  enzyme the lower is its
  efficiency.}\label{fig:Reaction-rate_vs_concentration}
\end{figure}

We are going to show that the increment of an enzyme is lower the
higher is its $K_M$ (remember that an high $K_M$ corresponds to a
low-efficiency of the enzyme, see also Fig. \ref{fig:Reaction-rate_vs_concentration}). Thus we can observe no significant change in the concentration of
a particular enzyme if its $K_M$ is very low.

If we look for example at glycolytic metabolism, it is possible to
observe no variation in glyceraldehyde 3-phosphate dehydrogenase
(GAPDH) concentration while triose-phosphate isomerase (TPI) increment
is significant. This could be due to the very low $K_M$ of GAPDH with
respect to the TPI one. Accordingly GAPDH can be considered an
``housekeeping'' enzyme.  

Fig. \ref{fig:Reaction-rate_vs_concentration} shows the
Michaelis-Menten curves of $2$ different enzymes $A$ and $B$,
respectively a slow and a fast enzyme. 

We want to show that (for example when a metabolic alteration occurs
inside the cell, since an higher ATP concentration is required) the
concentration of an enzyme $A$ with an high $K_M$ (less efficient)
varies more than the concentration of a more efficient enzyme $B$.

The enzyme rate-substrate concentration relationship reported in
\ref{fig:Reaction-rate_vs_concentration} refers to a single
enzyme. Although we are interested in cellular adaptation in terms of
variations in enzymes concentration. We know that the velocity of
conversion of substrates into products reflects the probability to
match the substrate inside the cell
\begin{align}\label{eq:velocity}
V_n=nV
\end{align}
with $n$ the copies of a single enzyme. Imagine to treat a cell with a
drug that inhibits glucose uptake (\eg glucocorticoids). In this case
there is less glucose in the cytoplasm so that the cell reacts
enhancing the production of glycolitic enzymes in order to increase
the probability to match their substrates.  Focus on two enzymes $A$
and $B$ following the Michaelis-Menten kinetics and assume for
simplicity the number of these two enzymes at time $t=0$ before the
treatment to be $n_{0}^A=n_0^B=1$.  After the treatment at time $t$ we
have $n_t^A=n^A$ and $n_t^B=n^B$ and according to
Eq. \eqref{eq:velocity} we have
\begin{align}
  \label{eq:after_treatment}
  V^{A}_{n^A}=n_AV^{A},\qquad   V^{B}_{n^B}=n_BV^{B}
\end{align}
It is realistic to suppose that they have different kinetics
properties and in particular suppose $A$ to be more efficient than $B$
\begin{align}
  \label{eq:relation}
  V^A>V^B
\end{align}
We are considering the alteration of a pathway with both the enzymes 
involved in this specific pathway, we can say that 
\begin{align}
  V^{A}_{n^A}=V^{B}_{n^B}  
\end{align}
and  from Eqs. \eqref{eq:after_treatment} and \eqref{eq:relation} we finally get
\begin{align}
n^{A}>n^{B}.  
\end{align}
We have shown that the increment of an enzyme with an high $K_M$ is
lower than the increase in a low-efficiency enzyme. 

But thinking about western blots another parameter to consider is the
antibody affinity for the target enzyme. Antibody affinity for the
ligand can be expressed in terms of the association constant $Y$ as:
\begin{align}
  Y=\frac{[Ab\times Lg]}{[Ab][Lg]}\qquad\text{at equilibrium}
\end{align}
where $[Ab]$ and $[Lg]$ denote respectively the antibody and the
ligand concentrations, while $[Ab\times Lg]$ denotes the concentration
of the antibody bind to its ligand. The enzyme concentration we observe with
western blot analysis $E_{obs}$ is related to the real concentration $E$
through this formula:
\begin{align}
E=\frac{1}{Y}\cdot E_{obs}
\end{align}
To compare the expression of an enzyme between control and treated
samples, it is not necessary to worry about $Y$. However, if we mind to
compare the modifications of different enzymes we have to consider
antibody affinity.  Combining the two parameters ($K_M$ and $Y$) we
obtain a more realistic indicator of adaptation $\Xi$  can be 
\begin{align}
  \Xi= \frac{1}{Y}\cdot\Delta E^1_{obs}\cdot K_{M_1}=\frac{1}{Y}\cdot\Delta E^2_{obs}\times K_{M_2}
\end{align}
where $1$ and $2$ label two different enzymes of the pathway under
consideration.  Finally, if we are studying a certain signal transduction pathways
component, we should take into account signal amplification
too. Signal amplification is not a small phenomenon. On the contrary,
one molecule leads to the activation of $10^6$.

\section{Game theory and signal transduction pathways dynamics}

\begin{figure}[h!]
\centering
\includegraphics[width=.8\textwidth]{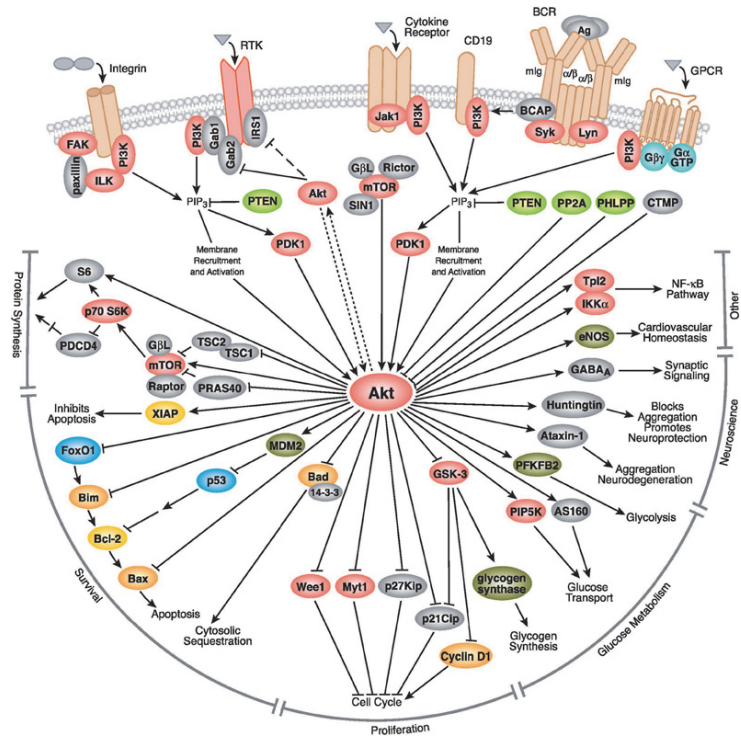}
\caption{PI3K/Akt Signaling.}\label{fig:PI3K-Akt_Signaling}
\end{figure}

Signal transduction pathways recover a crucial role in cellular
processes: they represent a connection between environmental
conditions and cellular reactions. It is well known that signals are
transduced from the cell surface to the cell nucleus by a series of
protein-protein interactions, phosphorylation reactions. Every signal
transduction pathway is composed by one receptor and some kinases that
bring the environmental signal to the nucleus. Usually, when the
ligand binds the receptor, it activates a kinase by prosphorylation,
the signal travel through the kinase and then the kinase activates the
next one in the chain. Different pathways are linked to create
biological networks.  In some networks a single protein is linked to
many others, as for Akt in Fig. \ref{fig:PI3K-Akt_Signaling}. Nevertheless it is improbable for a
protein to be optimized to interact with so many other proteins of
almost the same dimensions.

We can try to explain this considering inhibitions to be indirect
inhibitions. Indeed signal amplification leads to $10^6$ molecules per
cell, leaving no space for the amplification of a second pathway
not-related (with no links) to the first as cells dimension indicates
(a cell can contain about $10^8$ proteins). Hence it can not be active
more non-related pathways per time. This suggests that probably a
tissue is composed by a snap of an heterogeneous population.  In this
scenario it can also be possible to calculate the duration of a single
pathway activation knowing the lifetime of its components.  In \cite{chettaoui2006rewriting} is shown how the formalism of game
theory can be used to characterise biological cascades and gene
regulation.  In this paper we propose game theory as a tool for
studying signal transduction pathways. Indeed we introduce the
replicator dynamics differential equation used in evolutionary game
theory.

\begin{align}
\frac{\partial \ln{x_i(t)}}{\partial t}= x_i \left[u_i(x)-\sum_{j=1}^{n}x_j u_j(x)\right]
\end{align}
where $x_i$ is the strategy adopted by player $i$ and $u_i(x)$ is the
utility (or the fitness), while $\sum_{j=1}^{n}x_j u_j(x)$  is the average population fitness.

The replicator dynamics equation allows to calculate the population growing rate. 
In our case
\begin{align}
u(x)=\frac{1}{k_M}\cdot s
\end{align}
where $K_M$ is the Michaelis-Menten constant of the kinase, and￼ is the
amplification factor so that, when a signal arrives being a ligand, it
causes for example the activation and amplification of different Akt
downstream kinases witch fight to reach their targets. This
competition for the space indirectly inhibits the activation of other
pathway and it is responsible for the modulation in the cellular
response.

\section{Real Time PCR to detect cellular subpopulations}

Many progresses have been done since the introduction of Polymerase
Chain Reaction (PCR)
\cite{higuchi1993kinetic,heid1996real}. Microfluidics allows Real Time
PCR \cite{vandesompele2002accurate} which is quantitative in
opposition to classical PCR. The last innovation in this field is the
Droplet Digital \cite{hindson2011high}.  

In this Section we suggest an innovative application of Real Time PCR in
detecting cellular subpopulations in our sample.  Some tissues are
themselves heterogeneous populations (as in the case of muscles), in
other cases cells react in different ways to an injury. Think about
images for quantitative immunohistochemistry: they do not represent an
homogeneous pattern.
\begin{figure}[h!]
\centering
\includegraphics[width=.7\textwidth]{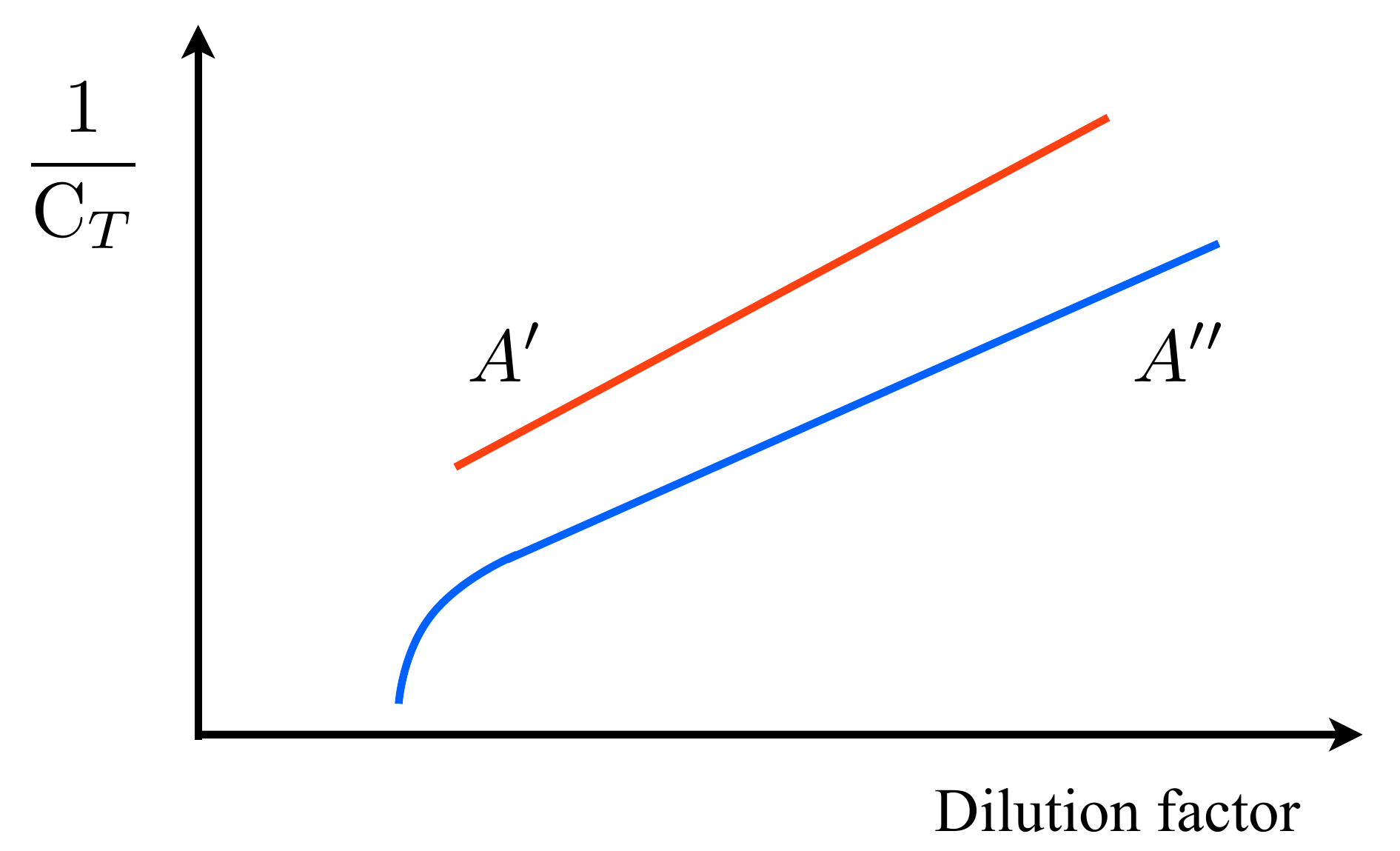}
\caption{In the picture $\mathrm{C}_T$ is the threshold cycle. This
  qualitative plot is to show that making serial dilution of cDNA
  extracted from the treated sample we obtain an exponential curve
  (the image represents the linear part only $A'$), we can say that
  there is no subpopulation while we can conclude the contrary if we
  get a Poissonian (scenario $A''$).}\label{fig:plot}
\end{figure}
This application of Real Time PCR can be useful overall for acute
phase studies.  Suppose to amplify one gene of your control and
treated sample and to obtain an induction of the ubiquitin ligases and
an induction of PGC-1alpha (oxidative metabolism regulator). A
possible interpretation of this data could be that some cells are
trying to adapt to the new condition while others are dying. Otherwise
imagine to obtain no significant increment. At this point one can
think that we need to increase the number of samples or that this gene
does not change under these conditions.  We can try to solve the
question with no more samples nor cDNA! We can make serial dilutions
of cDNA and if a cut-off appears with a nonlinear relationship, then
we can hypothesize the presence of a subpopulation in your
sample. Indeed, diluting cDNA we expect to find a progressive
decrease in cDNA concentration (above sensibility threshold),
depending on the amplification probability inside the well (sample
$A'$ in Fig. \ref{fig:plot}). But if there is a subpopulation inside
we also have to consider the probability to pipetting the fragments to
execute the Real Time PCR experiment (sample $A''$ in
Fig. \ref{fig:plot}).

\bibliographystyle{plain}
\bibliography{bibliography_vale}

\end{document}